\documentstyle[12pt]{article}
\begin{document}
\def\Tha{\Theta^{\alpha}}
\def\Thb{\Theta^{\beta}}
\def\da{{\dagger}}
\def\dda{{\ddagger}}
\def\tha{\theta^{\alpha}}
\def\thb{\theta^{\beta}}
\def\th+{\theta^+}
\def\th-{\theta^-}
\def\del{\partial}
\def\a{\alpha}
\def\e{\varepsilon}
\def\b{\beta}
\def\G{\Gamma}
\def\C{C_{\a\b}}
\def\cA{\cal{A}}
\def\os2{$osp(2,2)$}
\def\su{$su(2)$}
\def\be{\begin{equation}}
\def\ee{\end{equation}}
\def\bi{\bibitem}
\begin{titlepage}
\begin{flushright}
CERN-TH/96-63\\
UWThPh-21-1996\\
hep-th/9603071
\end{flushright}
\begin{center}
{\large\bf $N=2$ Superalgebra and Non-Commutative Geometry} \\
\vskip 1cm
{\bf H. Grosse\footnote{Participating in  Project No. P8916-PHY of the
`Fonds zur F\"orderung der wissenschaftlichen Forschung in
\"Osterreich'.}}\\
\vskip 0.3cm
{\it Institute for Theoretical Physics, University of Vienna,}\\
{\it Boltzmanngasse 5, A-1090 Vienna, Austria,}\\ \vskip 0.5cm
{\bf C. Klim\v c\'\i k} \\
\vskip 0.2cm
  {\it Theory Division CERN,\\ CH-1211 Geneva 23,
Switzerland} \\
\vskip 0.2cm {\small and} \\
\vskip 0.2cm
 {\bf P. Pre\v snajder}\\
\vskip 0.2cm
 {\it Dept. of Theoretical Physics, Comenius
University,} \\ {\it Mlynsk\'a dolina F1, SK-84215 Bratislava,
 Slovakia} \\
\end{center}
\vskip 0.5cm
\begin{abstract}
A  construction of supersymmetric field-theoretical models 
in non-commutative geometry is reviewed. The underlying 
superstructure of the models is encoded in $osp(2,2)$ superalgebra.  
\end{abstract}
\vskip 0.5cm
\begin{center}
{\it Based on a seminar presented by C. Klim\v c\'\i k  at LXIV. Les Houches 
session on `Quantum Symmetries',
August 1995}
\end{center}
\vskip 0.5cm
\noindent CERN-TH/96-63\\
March  1996
\end{titlepage}

\section{\bf Introduction}

Probably the most fruitful method for obtaining concrete nonperturbative 
quantitative
predictions for more or less realistic quantum field theories consists
in using some appropriate {\it finite dimensional} regularization of the
theory and then performing the path integral quantization.
As an example of such an approach we may mention the usual lattice
regularization (see \cite{Mon} for a review) which regularizes
the space itself and hence the algebra of functions on it.
 The power of the lattice
methods does not suffice to treat the supersymmetric theories however,
because the notion of the superspace
itself does not make a direct sense. Only the algebra of functions on the
superspace is well defined.

In this contribution, we wish to describe an alternative of the usual 
lattice regularization which would be well suited also for regularizing the 
supersymmetric theories. Our basic idea relies in regularizing directly
the superalgebra  of functions on the superspace rather than considering
  a regularization of the superspace itself. 
Moreover, we shall truncate the superalgebra of functions 
in a manner consistent with
supersymmetry and describe an appropriate modified differential and integral
calculus by using the methods of the non-commutative geometry \cite{Con},
 i.e. the generalization
of the ordinary differential geometry to non-commutative
rings of `functions'.
This calculus suits well for our basic purpose of  writing 
 down  regularized actions of supersymmetric field theories. 

 As the starting point of our
treatment we choose a 2d field theory on a truncated two-sphere\footnote{Also 
referred to as ``fuzzy'', ``non-commutative''
or ``quantum'' sphere in the literature \cite{Ber,Hop,Mad}.}\cite{Mad2,GKP1}.
 Apparently, the truncated sphere  was
introduced by Berezin, in 1975 \cite{Ber}, who quantized the
(symplectic) volume two-form on the ordinary two-sphere.
The concept was rediscovered  by Madore \cite{Mad}
(see also \cite{GroPre}) who has approached the structure
from the point of view of the non-commutative geometry.

 In what follows, we shall construct the supersymmetric extension of the
truncated sphere and build up a field theory on it \cite{GKP2}.
 The basic underlying
(super)structure is the Lie superalgebra $osp(2,2)$. The resulting 
theory is manifestly finite and $osp(2,1)$ supersymmetric. 
\section{Commutative supersphere}

Consider a three-dimensional superspace ${\bf SR^3}$ with coordinates
$x^i,\theta^{\alpha}$;
the supercoordinates are the \su ~Majorana spinors. Denote 
${\cal SB}$ the algebra of
 analytic functions on the superspace with the Grassmann coefficients in front
of the odd monomials in $\theta$.

The vector fields in ${\bf SR^3}$ generating \os2 ~super-rotations of
${\cal SB}$ are given by the explicit formulae \cite{GKP2}

\be v_+=-{1\over 2}\big(x^3\del_{\theta^-}-(x^1+ix^2)\del_{\theta^+}\big)+
{1\over 2}\big(-\theta^+\del_{x^3}-\theta^-(\del_{x^1}+i\del_{x^2})\big),\ee
\be v_-=-{1\over 2}\big(x^3\del_{\theta^+}+(x^1-ix^2)\del_{\theta^-}\big)+
{1\over 2}\big(\theta^-\del_{x^3}-\theta^+(\del_{x^1}-i\del_{x^2})\big),\ee
\be d_+=-{1\over 2}r(1+{2\over r^2}\theta^+ \th-)\del_- +{\th-\over 2r}R_+
-{\theta^+ \over 2r}(x^i\del_i-R_3),\ee
\be d_-={1\over 2}r(1+{2\over r^2}\theta^+ \th-)\del_+ +{\theta^+ \over 2r}R_-
-{\th-\over 2r}(x^i\del_i+R_3)\ee
\be \Gamma_{\infty}=\Big({\theta^+ x^3\over r}+{\th- x^+\over r}\Big)\del_+ +
\Big({\theta^+ x^-\over r}-{\th- x^3\over r}\Big)
\del_-  \equiv 2(\th- v_+ -\theta^+ v_-).\ee
\be r_+=x^3(\del_{x^1}+i\del_{x^2})-(x^1+ix^2)\del_{x^3}+\theta^+
\del_{\theta^-},\ee
\be r_-=-x^3(\del_{x^1}-i\del_{x^2})+(x^1-ix^2)\del_{x^3}+
\theta^-\del_{\theta^+},\ee
\be r_3=-ix^1\del_{x^2}+ix^2\del_{x^1}+{1\over 2}
(\theta^+\del_{\theta^+}-\theta^-\del_{\theta^-})\ee
and they obey the \os2 ~ Lie superalgebra graded commutation relations
\cite{ElgN,SNR}
\be [r_3,r_{\pm}]=\pm r_{\pm},\qquad [r_+,r_-]=2r_3,\ee
\be [r_3,v_{\pm}]=\pm{1\over 2}v_{\pm},\qquad [r_{\pm},v_{\pm}]=0,
\qquad [r_{\pm},v_{\mp}]=v_{\pm},\ee
\be\{v_{\pm},v_{\pm}\}=\pm{1\over 2}r_{\pm},\qquad \{v_{\pm},v_{\mp}\}=
-{1\over 2}r_3.\ee
\be [\G_{\infty},v_{\pm}]=d_{\pm},\qquad[\G_{\infty},d_{\pm}]=v_{\pm},
\qquad [\G_{\infty},r_i]=0,\ee
\be [r_3,d_{\pm}]=\pm {1\over 2}d_{\pm},\quad[r_{\pm},d_{\pm}]=0,
\quad [r_{\pm},d_{\mp}]=d_{\pm},\ee
\be \{d_{\pm},v_{\pm}\}=0,\qquad \{d_{\pm},v_{\mp}\}=
\pm{1\over 4}\G_{\infty},\ee
\be \{d_{\pm},d_{\pm}\}=\mp{1\over 2}r_{\pm},\qquad
\{d_{\pm},d_{\mp}\}={1\over 2}r_3.\ee

Note that the superfunction
\be \sum {x^i}^2+C_{\a\b}\tha\thb -\rho^2 , \quad  C=i\sigma^2\ee
is $osp(2,2)$-invariant. This means that the algebra 
${\cal SB}$, factorized by its ideal
${\cal SI}$
consisting of all functions of the  form $h(x^i,\tha)(\sum {x^i}^2+
C_{\a\b}\tha\thb -\rho^2)$, inheritates the  $osp(2,2)$ action\footnote{
The appearance of $r$ in Eqs. (3)-(5) may seem
awful because we have considered the ring of superanalytic functions
on $SR_3$. However, $r$ becomes
 harmless after the factorization by the ideal $SI$.}.
 We denote  the quotient as ${\cal SA}_{\infty}$ and refer to it
as to the algebra
of superfields on the supersphere. 

An \os2 ~invariant
 inner product of two elements $\Phi_1,\Phi_2$ of
 ${\cal SA}_{\infty}$ is given by\footnote{The normalization ensures that
the norm of the unit
element of ${\cal SA}_{\infty}$ is 1. The inner product is  supersymmetric
but it is not positive definite. However, such a property of the product is
not needed for our purposes.}
\be(\Phi_1,\Phi_2)_{\infty}\equiv{\rho\over 2\pi}\int_{R^3}d^3 x^i
d\theta^+ d\theta^-\delta
({x^i}^2+C_{\a\b}\tha\thb -\rho^2)\Phi_1^{\ddagger}(x^i,\tha)
\Phi_2(x^i,\tha),\ee
Here $\Phi_1(x^i,\tha),\Phi_2(x^i,\tha)\in{\cal SB}$ are some representatives
of $\Phi_1$ and $\Phi_2$ and the (graded) involution \cite{SNR2,SNR} is
defined
by
\be {\theta^+}^\dda=\theta^-,~{\theta^-}^\dda=-\theta^+,~
{}~(AB)^\dda=(-1)^{degA~deg B}B^\dda A^\dda .\ee The algebra
${\cal SA}_{\infty}$ is obviously generated by (the equivalence classes)
$x^i~ (i=1,2,3)$ and $\tha~ (\a=+,-)$ which (anti)commute
with each other under the usual pointwise multiplication, i.e.
\be x^i x^j-x^j x^i=x^i\tha-\tha x^i=\tha\thb+\thb\tha=0.\ee Their norms are
given by
\be \vert\vert x^i\vert\vert^2_{\infty}=\vert\vert \tha\vert\vert^2_{\infty}=
\rho^2.\ee

As  is well known \cite{SNR}, the typical irreducible representations of
\os2 ~ consist of quadruplets of the \su ~
irreducible representations $j\oplus j-{1\over 2}\oplus j-{1\over 2}\oplus
j-1$. The number $j$ is an integer
or a half-integer, and it is referred to as the \os2 ~superspin.
The supermultiplet with the superspin $1$ can be conveniently constructed,
applying subsequently the lowering operators $v_-$ and $d_-$ on the highest
weight vector $x^+$.
 Supermultiplets with higher superspins can be obtained
in the same way, starting with the highest weight vectors ${x^+}^l$.
Thus
 the full decomposition of ${\cal SA}_{\infty}$ into the irreducible
representations of \os2 ~ can be written as  the infinite
direct sum
\be {\cal SA}_{\infty}=0+1+2+\dots,\ee
where the integers  denote the \os2 ~superspins of the
representations.
 From the point of view of the \su ~representations,
the algebra of the superfields consists of two copies of the standard
algebra of scalar fields on the ordinary  sphere $S^2$
and one copy of the spinor bundle  on $S^2$.
 Note that the generators of ${\cal SA}_{\infty}$
fulfil the obvious relation
\be {x^i}^2+C_{\a\b}\tha\thb =\rho^2.\ee

Now we turn to the problem of formulating supersymmetric field theories
on the sphere. We restrict our attention to the case of $N=1$ supersymmetry,
hence we look for an $osp(2,1)$-invariant action for the superscalar
field from ${\cal SA}_{\infty}$. It is easy to check that the operator
\be C_{\a\b}d_{\a}d_{\b}+{1\over 4}\G_{\infty}^2\ee
is invariant with respect to $osp(2,1)$ supersymmetry generated by $r_i$
and $v_{\pm}$. We observe that the additional odd $osp(2,2)$ 
generators $d_{\a}$ play simply the role of the supersymmetric covariant
derivative in the following $osp(2,1)$ invariant action
$$ S=(\Phi,C_{\a\b}d_{\a}d_{\b}\Phi)_{\infty}+
{1\over 4}(\Phi,\G_{\infty}^2\Phi)_{\infty}\equiv $$
$$\equiv{\rho\over 2\pi}\int_{R^3}d^3 x^i d\theta^+ d\th-
\delta({x^i}^2+C_{\a\b}\tha\thb -\rho^2)\Phi(x^i,\tha)
(\C d_{\a}d_{\b} +{1\over 4}\G_{\infty}^2)\Phi(x^i,\tha),$$
where $\Phi$ is a real superfield, i.e. $\Phi^{\dda}=\Phi$.
Note that the additional bosonic $osp(2,2)$ generator $\G_{\infty}$
also appears in the action. 

Consider  the variation of the real superfield $\Phi$
\be \delta\Phi=i\e_{\a}v_{\a}\Phi,\ee
which preserves the reality condition. Using the fact
that $\e_{\a}v_{\a}$ commutes with the operator (23), the
supersymmetry of the action $S$ easily follows.

Now we should discuss
 the crucial importance of $\G_{\infty}$ in our formalism.
Indeed, after the obvious
identification of the $\theta$-linear part of the
superfield
\be \Phi(x^i,\tha)=\phi(x^i)+\psi_{\a}\tha +(F+{x^i\over r^2}\del_i
\phi)\theta^+ \th- \ee
with spinors on the sphere we observe that $\G_{\infty}$ is simply 
the grading operator of the spinor bundle. It anticommutes with
the operator (23) restricted to fermions or, in other words, with the
standard round Dirac operator on the sphere. Indeed,
it is straightforward to work out the action  in the
two-dimensional component language. It reads
\be S={1\over 4\pi}\int d\Omega \Big(-{1\over 2}\phi
\bigtriangleup_{\Omega}\phi+
{1\over 2}
\rho^4 F^2 -{1\over 2}\psi^{\da}\rho^3 D_{\Omega}\psi\Big),\ee
where $D_{\Omega}$ is the round  Dirac operator on $S^2$. 
We recognize in  this expression
the standard free supersymmetric action in two
dimensions.

The fact that the chirality operator $\G_{\infty}$ is 
at the same time the $osp(2,2)$-generator is very important because
any $osp(2,2)$-covariant regularization of the SUSY field theories will
automatically be also chiral. Needless to say, particularly this aspect
appears to be very promising from the point of view of higher dimensional
generalizations of our formalism.

We conclude this section by noting that by adding
 a (real) superpotential $W(\Phi)$ we may write down a
supersymmetric action with the interaction term. It reads
\be S_{\infty}=\Big(\Phi,\big(\C d_{\a}d_{\b}+
{1\over 4}\G_{\infty}^2\big)\Phi\Big)_{\infty}+
(1,W(\Phi))_{\infty}.\ee

\section{Non-commutative supersphere}

We define the non-commutative supersphere ${\cal{SA}}_j$ 
by means of the truncation
of the expansion (21). This means that the \os2 ~ decomposition
of the superalgebra of superfunctions terminates at the value
of the maximal \os2 ~ superspin $j$ which will play the role of cut-off
in our regularization. Hence 
\be {{\cal{SA}}_j}=0+1+\dots+j,\qquad j\in{\bf Z}\ee
as the vector space. We have to furnish this linear space
with an associative product and an inner product,
 which in the limit $j\to\infty$,
give the standard products in ${\cal{SA}}_{\infty}$.

In order to do this consider the space ${\cal{L}}(j/2,j/2)$ of linear
operators
from the representation space of the $osp(2,1)$ ~irreducible representation
with the
$osp(2,1)$ superspin $j/2$ into itself. (Note that the $osp(2,1)$
irreducible representation
 with the $osp(2,1)$ superspin $j$ has the \su ~content $j
\oplus j-{1\over 2}$ \cite{SNR}.)  The action of the superalgebra
\os2~itself on
the $j/2$ representation space\footnote{The so-called non-typical irreducible
representation
of \os2 ~ \cite{SNR,Mar}  is at the same time  the $osp(2,1)$
irreducible representation with the
$osp(2,1)$ superspin $j/2$.} is described by the operators $R_i,V_{\a},D_{\a},
\gamma\in{\cal{L}}(j/2,j/2)$ given by \cite{PR}:
\be R_i=
\left(\matrix{R_i^{j\over 2}&0\cr 0&R_i^{{j\over 2}-{1\over 2}}}\right),
\qquad \gamma=\left(\matrix{-j~Id&0\cr 0& -(j+1)Id}\right).\ee
\be
 V_{\a}=\left(\matrix{0&V_{\a}^{{j\over 2},{j\over 2}-{1\over 2}}\cr
V_{\a}^{{j\over 2}-{1\over 2},{j\over 2}}&0}\right),
\qquad D_{\a}=\left(\matrix{0&V_{\a}^{{j\over 2},{j\over 2}-{1\over 2}}\cr -
V_{\a}^{{j\over 2}-{1\over 2},{j\over 2}}&0}\right)
,\ee

where
\be \langle l,l_3+1\vert R_+^l\vert l,l_3\rangle=\sqrt{(l-l_3)(l+l_3+1)},\ee
\be \langle l,l_3-1\vert R_-^l\vert l,l_3\rangle=\sqrt{(l+l_3)(l-l_3+1)},\ee
\be \langle l,l_3\vert R_3^l\vert l,l_3\rangle=l_3,\ee
\be \langle l_3+{1\over 2} \vert V_+^{{j\over 2},{j\over 2}-{1\over 2}}
\vert l_3\rangle=-{1\over 2}\sqrt{{j\over 2}+l_3+{1\over 2}},\ee
\be \langle l_3-{1\over 2} \vert V_-^{{j\over 2},{j\over 2}-{1\over 2}}\vert
l_3\rangle=-{1\over 2}\sqrt{{j\over 2}-l_3+{1\over 2}},\ee
\be \langle l_3+{1\over 2} \vert V_+^{{j\over 2}-{1\over 2},{j\over 2}}\vert
l_3\rangle=-{1\over 2}\sqrt{{j\over 2}-l_3},\ee
\be \langle l_3-{1\over 2} \vert V_-^{{j\over 2}-{1\over 2},{j\over 2}}\vert
l_3\rangle={1\over 2}\sqrt{{j\over 2}+l_3}.\ee
 Every  $\Phi\in{\cal{L}}(j/2,j/2)$ can be written as a matrix
\be \Phi=\left(\matrix{\phi_R &\psi_R\cr \psi_L &\phi_L}\right),\ee
where $\phi_R$ and $\phi_L$ are square $(j+1)\times (j+1)$ and $j\times j$
matrices, respectively, and $\psi_R$ and $\psi_L$ are respectively rectangular
$(j+1)\times j$ and $j\times(j+1)$ matrices. A fermionic  element is
given by
a supermatrix with vanishing diagonal blocks, and a bosonic element by one
with vanishing off-diagonal blocks.
Clearly, \os2 \ superalgebra
acts on ${\cal{L}}(j/2,j/2)$ by the superadjoint action
\be {\cal R}_i\Phi\equiv [R_i,\Phi],\qquad \Gamma\Phi\equiv [\gamma,\Phi].\ee
\be {\cal V}_{\a}\Phi_{even}\equiv [V_{\a},\Phi_{even}],
\qquad {\cal V}_{\a}\Phi_{odd}\equiv \{V_{\a},\Phi_{odd}\}.\ee
\be {\cal D}_{\a}\Phi_{even}\equiv [D_{\a},\Phi_{even}],
\qquad {\cal D}_{\a}\Phi_{odd}\equiv \{D_{\a},\Phi_{odd}\}.\ee

 This `superadjoint'
 representation
is reducible and, in the spirit of Refs.\cite{SNR, Mar}, it is easy to work
out its decomposition into \os2 ~ irreducible representations
\be {\cal{L}}(j/2,j/2)=0+1 +\dots +j = {{\cal {SA}}_j}.\ee
 The associative product in ${\cal{L}}(j/2,j/2)$ is defined as the
composition of operators,  and
the \os2 ~invariant inner product on ${\cal{L}}(j/2,j/2)$ is defined
by
\be (\Phi_1,\Phi_2)_j\equiv {\rm STr}(\Phi_1^\dda,\Phi_2),
\qquad \Phi_1,\Phi_2\in {\cal{L}}(j/2,j/2).
\ee
  The supertrace STr is defined as usual
\be {\rm STr}\Phi\equiv {\rm Tr}\phi_R -{\rm Tr}\phi_L\ee
and the graded involution $\dda$ as \cite{SNR2}
\be \Phi^\dda\equiv
\left(\matrix{\phi_R^\da &\mp\psi_L^\da\cr\pm\psi_R^\da &\phi_L^\da}\right).\ee
$\da$ means the standard Hermitian conjugation of a matrix and the upper
(lower) sign refers to the case when the entries consists of odd (even)
elements
of a Grassmann algebra.
Note that
\be R_i^\dda=R_i,\qquad V_+^\dda=V_-,\quad V_-^\dda=-V_+.\ee
Now we  identify   ${\cal SA}_j$ with even  elements of ${\cal{L}}(j/2,j/2)$,
which means that the entries of the diagonal matrices are
commuting numbers and those of the off-diagonal matrices in turn 
are  Grassmann variables.

The $osp(2,1)$ Casimir reads
\be R_i^2+C_{\a\b}V_{\a}V_{\b} \ee
and its value in the $j/2$ representation is $j(2j+1)/4$.
We may renormalize the $osp(2,1)$ generators $R_i$ and $V_{\a}$ essentially
by the square-root of the value of the Casimir; the renormalized generators
we denote as $X^i_j$ and $\Theta^{\a}_j$ and $j$ refers to the superspin.
 More precisely, we require that in each irreducible representation the
new generators fulfil the defining equation of the supersphere:
 \be {X_j^i}^2+C_{\a\b}\Tha_j\Thb_j=\rho^2.\ee
It can be shown \cite{GKP2} that the objects $X^i_{\infty}$ and 
$\Tha_{\infty}$ gives the standard commutative generators $x^i$ and
 $\theta^{\a}$ of the previous section. The simplest test of consistence
of this statement may be provided by looking at the commutation relations
of the renormalized generators:
\be [X_j^m,X_j^n]=
i{\rho\over\sqrt{{j\over 2}({j\over 2}+{1\over 2})}}\epsilon_{mnp}X_j^p ,
\qquad\ee
\be [X_j^i,\Theta_j^{\a}]=
{\rho\over2\sqrt{{j\over 2}({j\over 2}+{1\over 2})}}{\sigma^i}^{\b\a}
\Theta_j^{\b},\ee
\be \{\Theta_j^{\a},\Theta_j^{\b}\}={\rho\over2\sqrt{{j\over 2}({j\over 2}+
{1\over 2})}}
(C\sigma^i)^{\a\b}X_j^i.\ee

Consider now the variation of a superfield $\Phi$
\be \delta\Phi=i(\epsilon_+ {\cal V}_+ +\epsilon_- {\cal V}_-)\Phi,\ee
where $\epsilon_{\a}$ is given by
\be \epsilon_{\a}=\left(\matrix{\e_{\a}&0\cr 0&-\e_{\a}}\right)\ee
and $\e_{\a}$ are the usual Grassmann variables with the involution properties
\be \e^{\dda}_+=\e_-,\quad \e^{\dda}_-=-\e_+ . \ee

Because the inner product (43) is $osp(2,2)$ invariant, we can 
easily demonstrate the invariance of the following
action (defined on the truncated supersphere) under the $osp(2,1)$ variation
(52):
\be S_j=\Big(\Phi,\big(\C {\cal D}_{\a}{\cal D}_{\b}+{1\over 4}\G^2\big)
\Phi\Big)_j+
(1,W(\Phi))_j.\ee
The inner product appearing here was defined in (43), the operators 
${\cal D}_{\a}$ and $\G$ in (39) - (41)  and the field $\Phi$ in (38).
We may also write down the regularized action
for the supersymmetric $\sigma$-models describing the superstring
propagation in curved backgrounds:

\be S_j=({\cal D}_+\Phi^A,g_{AB}(\Phi){\cal D}_+\Phi^B)_j
+({\cal D}_-\Phi^A,g_{AB}(\Phi){\cal D}_-\Phi^B)_j+{1\over 4}(\G\Phi^A,
g_{AB}(\Phi)\G\Phi^B)_j,\ee
where $g_{AB}(\Phi)$ denotes the metric on the target space manifold
in which superstring propagates.
The $osp(2,1)$ supersymmetry and the commutative limit are obvious.
 
The regularized actions (55) and (56) can
be used as the base for the path integral quantization, manifestly
preserving supersymmetry and still involving only the finite number of
degrees of freedom. Particularly this aspect of our approach seems
to be very promising both in comparison with  lattice physics
and also in general. 
\section{Outlook}

There are at least two obvious ways how to proceed
in order to make the method better elaborated and more universal.
The first one  would consists in introducing supergauge fields
in the formalism and the second one in supersymmetrizing the
truncated four-dimensional  bosonic sphere \cite{GKP4}.
We believe that a  success of the four-dimensional program would  
result in immediate realistic physical applications.


\newpage
\begin{thebibliography}{99}
\bibitem{Mon} I. Montvay and G. Muenster, {\it
Quantum fields on a lattice}, (Cambridge Univ. Press, Cambridge, 1994), and
references therein
\bibitem{Con}A. Connes, {\it Noncommutative Geometry}, (Academic Press, London,
1994)
\bibitem{Ber} F.A. Berezin, {\it Commun. Math. Phys.} {\bf 40} (1975) 153
\bibitem{Hop} J. Hoppe, {\rm MIT PhD Thesis}, 1982 and {\it Elem. Part. Res. J.
(Kyoto)} {\bf 80} (1989) 145
\bibitem{Mad} J. Madore, {\it J. Math. Phys.} {\bf 32} (1991) 332 
\bibitem{Mad2} J. Madore,
{\it Class. Quantum
Grav.} {\bf 9} (1992) 69
\bibitem{GKP1}H. Grosse, C. Klim\v c\'{\i}k and P. Pre\v snajder,
{\it Towards
Finite Quantum Field Theory in Non-commutative Geometry}, preprint
CERN-TH/95-138, UWThPh-19-1995, hep-th/9505175
\bibitem{GroPre} H. Grosse and P. Pre\v snajder, {\it Lett. Math. Phys.}
{\bf 28}
(1993) 239
\bibitem{GKP2} H. Grosse, C. Klim\v c\'\i k and P. Pre\v snajder,
 {\it Field Theory on a Supersymmetric Lattice},
 preprint CERN-TH/95-195, hep-th/9507074
\bibitem{ElgN} A. El Gradechi and L. Nieto, {\it Supercoherent States,
 Super-K\"ahler Geometry and Geometric Quantization}, Montreal preprint CRM-1876, 1994,
hep-th/9403109
\bibitem{SNR} M. Scheunert, W. Nahm and V. Rittenberg, {\it J. Math. Phys.}
{\bf 18} (1977) 154
\bibitem{SNR2} M. Scheunert, W. Nahm and V. Rittenberg, {\it J. Math. Phys.}
{\bf 18} (1977) 146
\bibitem{Mar} M. Marcu, {\it J. Math. Phys.} {\bf 21} (1980) 1284
\bibitem{PR} A. Pais and V. Rittenberg, {\it J. Math. Phys.} {\bf 16} (1975)
2062
\bibitem{GKP4} H. Grosse, C. Klim\v c\'\i k and P. Pre\v snajder,
 {\it On Finite 4d-Quantum Field Theory in Non-Commutative Geometry},
 preprint CERN-TH/96-51, hep-th/9602115
\end{thebibliography}
\end{document}